\documentclass[pss]{wiley2sp} 
\usepackage{amsmath,bm}

\newcommand{\kk}{{\bm k}}
\newcommand{\rl}{\rangle\!\langle}

\newcommand{\hh}{{\cal G}}
\newcommand{\rr}{{\bf r}}

\begin{document}

\title{Theoretical study of phonon-assisted singlet-singlet relaxation in two-electron semiconductor quantum dot molecules}

\titlerunning{Theoretical study of phonon-assisted singlet-singlet relaxation}

\author{  Anna Grodecka\textsuperscript{\textsf{\bfseries 1,\Ast}},
  Pawe{\l} Machnikowski\textsuperscript{\textsf{\bfseries 2}},
  Jens F{\"o}rstner\textsuperscript{\textsf{\bfseries 1}}}

\authorrunning{A. Grodecka et al.}

\mail{e-mail \textsf{anna.grodecka@uni-paderborn.de}, Phone
  +49-5251-602327, Fax +49-5251-603435}
  
\institute{ 
  \textsuperscript{1}\,Computational Nanophotonics Group, Theoretical Physics, 
University Paderborn, 33098 Paderborn, Germany\\
  \textsuperscript{2}\,Institute of Physics, Wroc{\l}aw University of Technology,
50-370 Wroc{\l}aw, Poland}

\received{XXXX, revised XXXX, accepted XXXX}
\published{XXXX}

\pacs{73.21.La, 03.65.Yz, 63.20.kd}

\abstract{
\abstcol{ Phonon-assisted singlet-singlet relaxation in semiconductor 
quantum dot molecules is studied theoretically. Laterally coupled quantum dot structures 
doped with two electrons are considered. We take into account interaction with acoustic phonon modes via deformation potential and piezoelectric coupling. We show that piezoelectric mechanism
for the considered system is of great importance and for some ranges of quantum dot molecule
parameters is the dominant contribution to relaxation. }
{It is shown that the phonon-assisted tunneling rates reach much higher values (up to 160~ns$^{-1}$ even at zero temperature) in comparison with other decoherence processes like spin-orbit coupling ($\sim$~0.01~ns$^{-1}$).
The influence of Coulomb interaction is discussed and its consequences are indicated.
We calculate the relaxation rates for GaAs quantum dot molecules and study the dependence on quantum dot size, distance and offset between the constituent quantum dots. In addition the temperature dependence of the tunneling rates is analyzed.}
}

\maketitle 

\section{Introduction}
Quantum dot molecules (QDMs) have attracted much interest in experimental \cite{bayer01,petta05} and theoretical \cite{rozbicki08} investigations due to their proposed application in various implementations of quantum information processing schemes. 
In particular, two-electron spin states in coupled quantum dots (QDs) have been employed 
in a number of realizations of quantum gates \cite{loss98,barrett06,hanson07}. 
Since many of them make use of or want to avoid electron tunneling processes, 
their timescales are of primary importance.

Electrons confined in semiconductor QDs interact with the phonon environment, which leads to loss of coherence. Tunneling coupling assisted by phonon degrees of freedom opens an additional decoherence channel. In the case of doubly doped structures, the Coulomb interaction plays also an important role. 

In this contribution, we study phonon-assisted singlet-singlet relaxation in laterally coupled GaAs QDs. QDMs doped with two electrons are analyzed and Coulomb interaction between them is included. Interaction with acoustic phonons via deformation potential and piezoelectric coupling is taken into consideration.
We show that the rates of phonon-assisted singlet-singlet relaxation are high (in the picosecond range) 
and strongly depend on the system parameters and temperature.

\section{The model}

A QDM doped with two electrons is considered. The Hamiltonian of the electron subsystem reads
\begin{equation}\label{eq:ham}
H_{\rm e} = \frac{\hbar^{2}}{2 m^{*}}\left( \nabla_{\rm a}^{2} +
\nabla_{\rm b}^{2} \right) + U(\rr_{\rm a}) + U(\rr_{\rm b})+
V(\rr_{\rm a},\rr_{\rm b}),
\end{equation}
where $m^{*} = 0.07 m_{0}$ is the effective mass of an electron in GaAs and 
$V (\rr_{\rm a},\rr_{\rm b}) = e^{2}/(4 \pi \varepsilon_{0} \varepsilon_{\rm r}
|\rr_{\rm a}-\rr_{\rm b}|)$ is the Coulomb interaction between the electrons.
Here, $e$ denotes electron charge, $\varepsilon_{0}$ is the vacuum
dielectric constant, and $\varepsilon_{\rm r}$ is the static relative
dielectric constant. The confinement potential $U(\rr_{\rm a/b})$ 
for two electrons referred to as `a' and `b' is assumed to be separable: 
$U(\rr) = \frac{1}{2} m^{*} \omega_{z}^{2} \; z^{2} 
+ \frac{1}{2} m^{*} \omega_{y}^{2} \; y^{2} + U(x)$, where the lateral potential describing the double QD
is chosen in the form
\begin{equation}
U(x) = - U_{0} e^{ - \frac{1}{2} \left( \frac{x - d/2}{a} \right)^{2} }
 - (U_{0} + \Delta U) e^{ - \frac{1}{2} \left( \frac{x + d/2}{a} \right)^{2}}.
\end{equation}
Here, $d$ denotes the distance between QDs, $a$ is the width of wave function in the $x$ and $y$ direction, $U_{0}$ and $U_{0} + \Delta U$ are QDs confinement depths and the difference
$\Delta U$ between them is referred to as the \textit{offset}.
The dynamics in the~$y$ and~$z$ directions is restricted to the ground states described by Gaussian wave functions. 

Two-particle spin-singlet states, labeled as $|0\rangle$ and $|1\rangle$, are constructed from the numerically obtained single particle wave functions, where the Coulomb interaction between electrons is also included \cite{grodecka08}.

The Hamiltonian describing the interaction of the electron singlet states with phonons reads
\begin{equation}
H_{\rm int} = |0 \rl 1| \sum_{s,\kk} G_{s}(\kk)
\left(b_{s,\kk}^{\phantom{\dag}} + b_{s,-\kk}^{\dag} \right) + \mathrm{H.c.},
\end{equation}
where $\kk$ is the phonon wave vector, $b_{s,\kk}^{\dag}$ and $b_{s,\kk}^{\phantom{\dag}}$ 
are phonon creation and annihilation operators, respectively. $G_{s}(\kk)$ are the two-electron coupling constants for different phonon branches $s$. We take into account interaction via deformation potential and piezoelectric coupling with respective coupling elements:
\begin{eqnarray}
G^{\rm DP}_{{\rm l}}(\kk) & = & \sqrt{\frac{\hbar k}{2 \rho V c_{\rm l}}}
D_{\rm e} \hh (\kk),\\
G^{\rm PE}_{s}(\kk) & = & -i \sqrt{\frac{\hbar}{2 \rho V c_s k}}
\frac{d_{\rm P} e}{\varepsilon_{0}\varepsilon_{\rm r}} M_s (\hat\kk) \hh (\kk).
\end{eqnarray}
Here, $\rho$ denotes the crystal density, $V$ is the normalization volume
of the phonon modes, $c_{s}$ is the speed of sound (longitudinal or transverse), 
$D_{\rm e}$ is the deformation potential constant for electrons 
and $d_{\rm P}$ is the piezoelectric constant. The form factor $\hh (\kk)$ reflects 
the geometrical properties of the wave functions and $M_s (\hat\kk)$ is a function depending only on $\kk$ orientation (for details see \cite{grodecka08}).

We describe the properties of the phonon reservoir by means of the phonon spectral density
\begin{eqnarray}
  \lefteqn{R (\omega) =} \\ \nonumber
  && \frac{1}{\hbar^{2}} |n_{\kk}+1| \sum_{s,\kk} |G_{s}(\kk)|^{2}  \left[
\delta(\omega-\omega_{s,\kk}) + \delta(\omega+\omega_{s,\kk}) \right],
  \end{eqnarray}
where $n_{\kk}$ is the Bose distribution function.

The phonon-assisted tunneling rates are calculated within the Markovian approach.
To this end, we employ the Fermi's golden rule and obtain the rate in the form
\begin{equation}
w = 2\pi R\left(\frac{\Delta E}{\hbar}\right).
\end{equation}
The relaxation rates are proportional to the phonon spectral density at the frequency corresponding to the splitting energy $\Delta E$, which is the difference between the two singlet states energies.

\section{Results}

In this section, we present and discuss the results for phonon-assisted tunneling rates in GaAs quantum dot molecule.
In the first part, relaxation rates are calculated at $T=0$~K, at the end the temperature dependence is shown.

\begin{figure*}[h]%
{\includegraphics*[width=.3\textwidth,height=1.0cm]{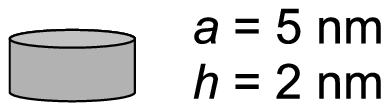}}\hfill
{\includegraphics*[width=.3\textwidth,height=1.0cm]{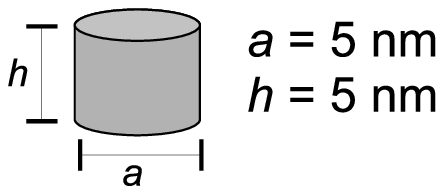}}\hfill
{\includegraphics*[width=.3\textwidth,height=1.0cm]{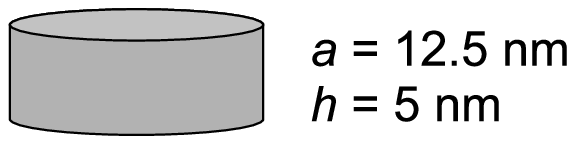}}\hfill
{\includegraphics*[width=.33\textwidth,height=4.0cm]{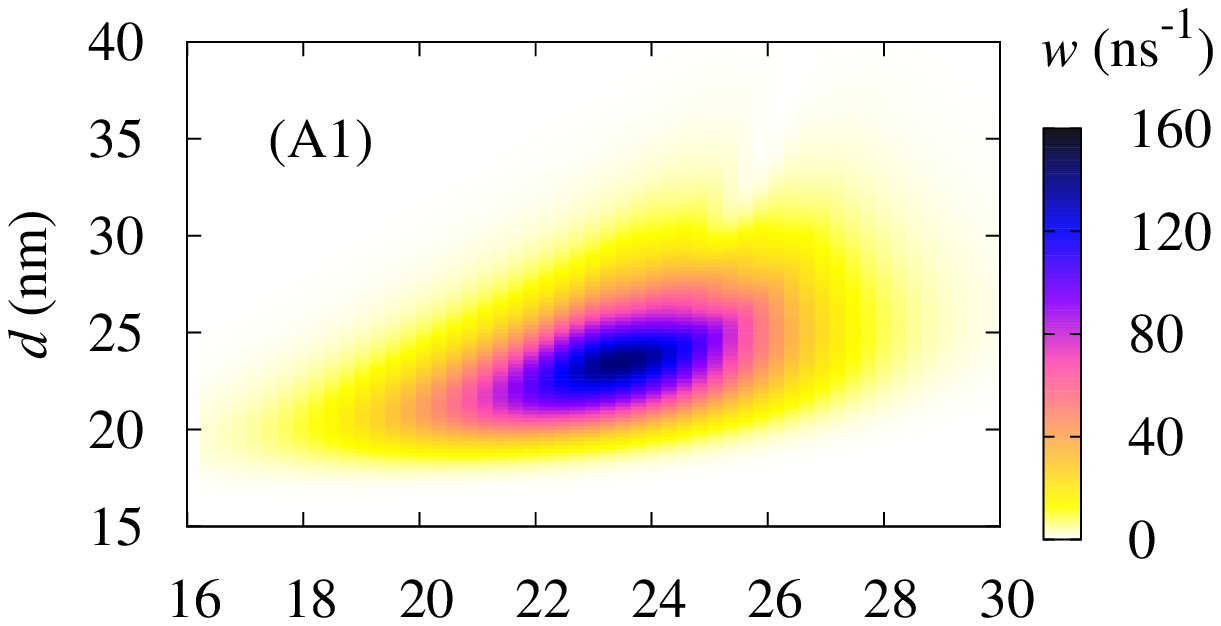}}\hfill
{\includegraphics*[width=.33\textwidth,height=4.0cm]{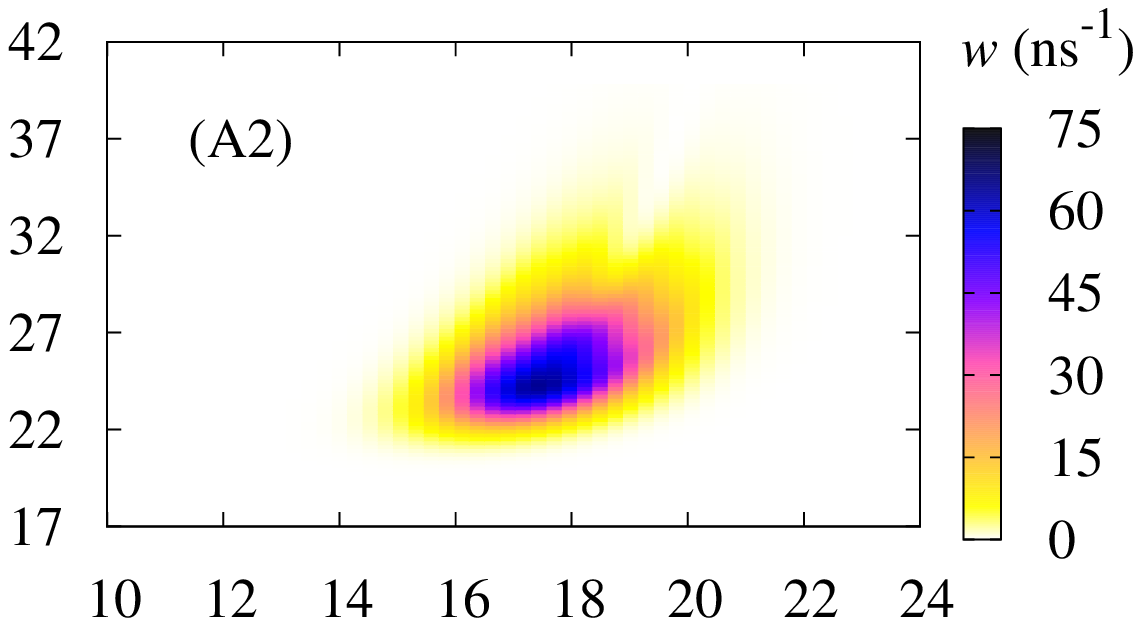}}\hfill
{\includegraphics*[width=.33\textwidth,height=4.0cm]{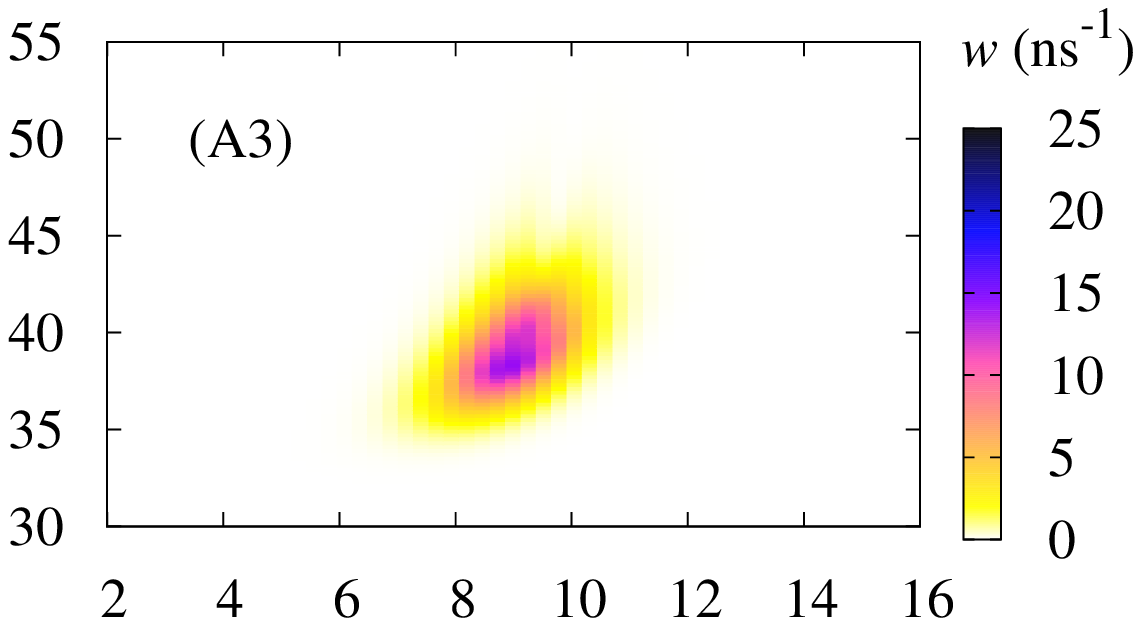}}\hfill
{\includegraphics*[width=.33\textwidth,height=4.0cm]{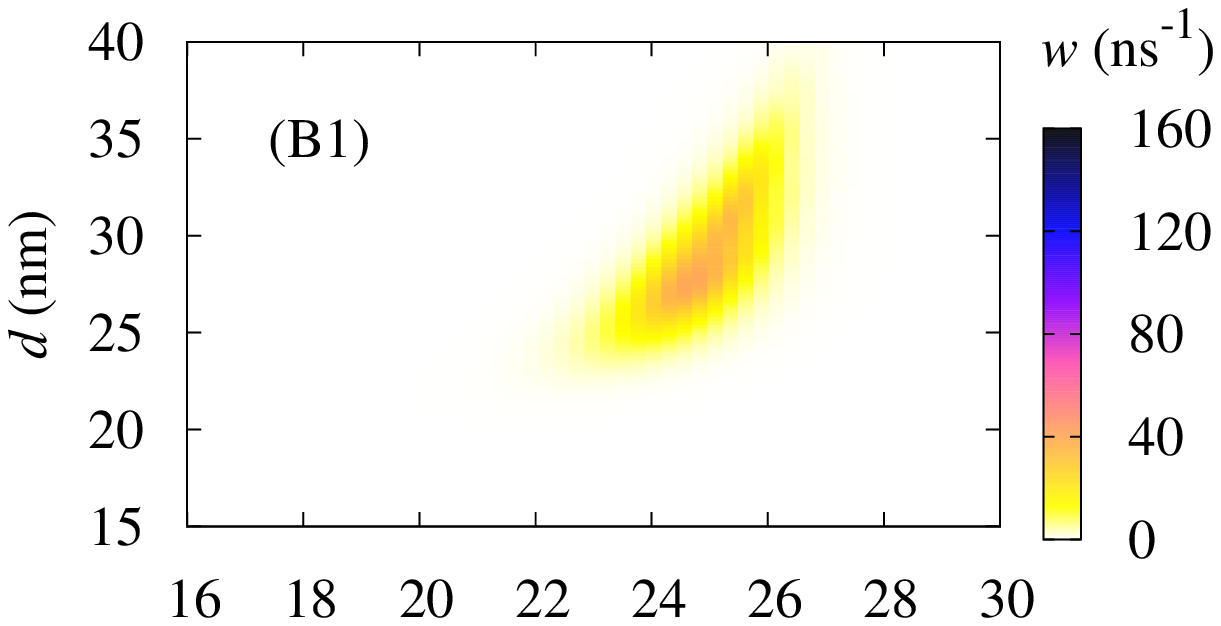}}\hfill
{\includegraphics*[width=.33\textwidth,height=4.0cm]{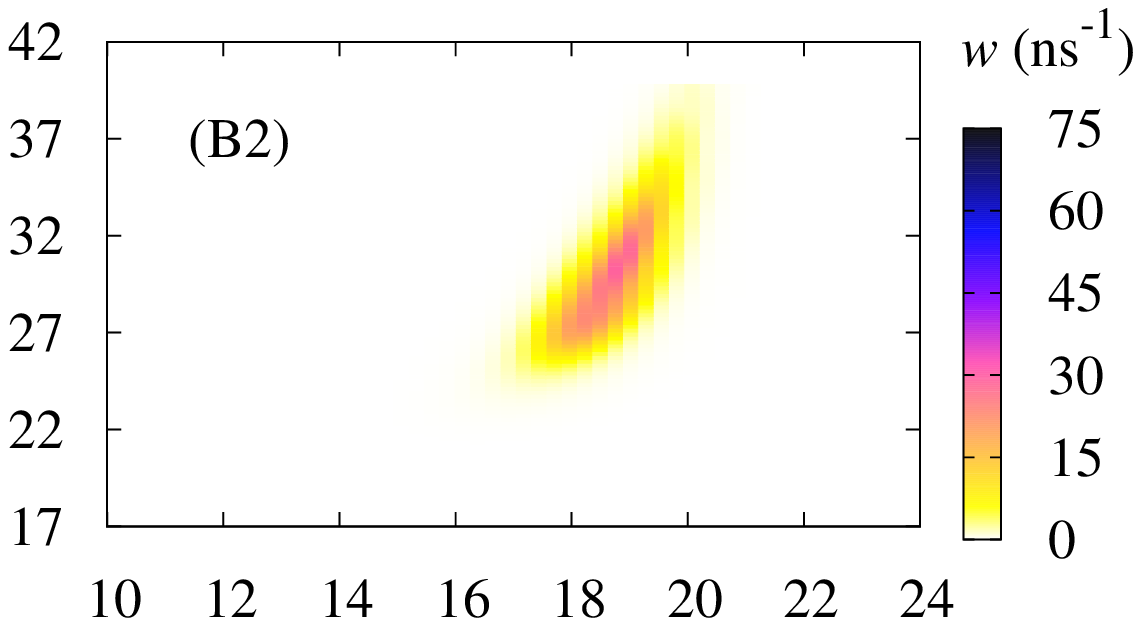}}\hfill
{\includegraphics*[width=.33\textwidth,height=4.0cm]{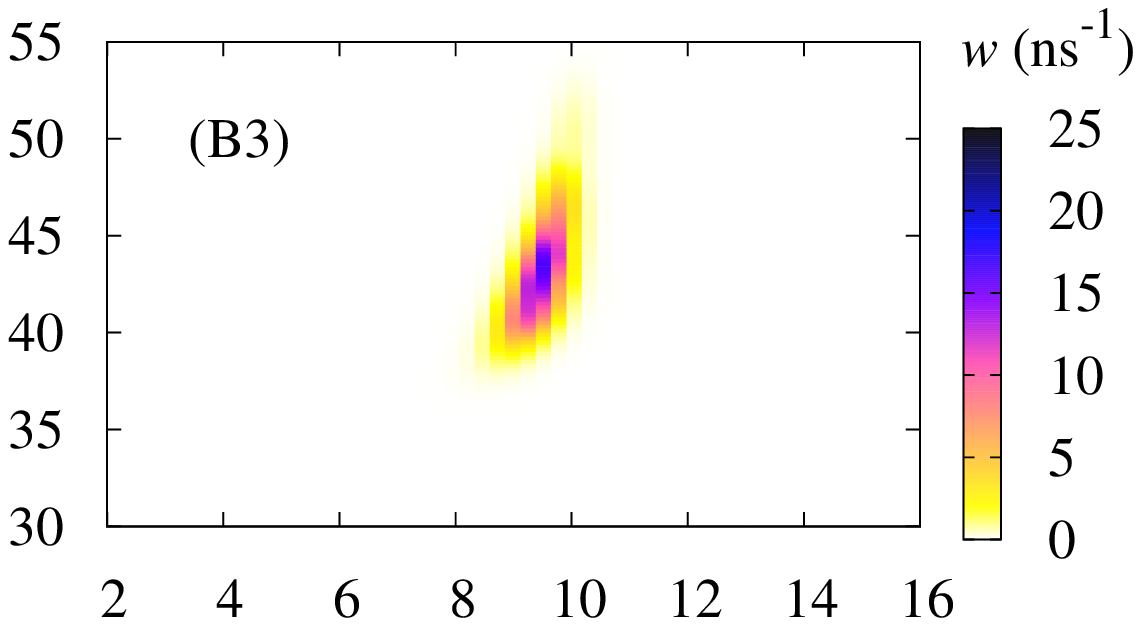}}\hfill
{\includegraphics*[width=.33\textwidth,height=4.0cm]{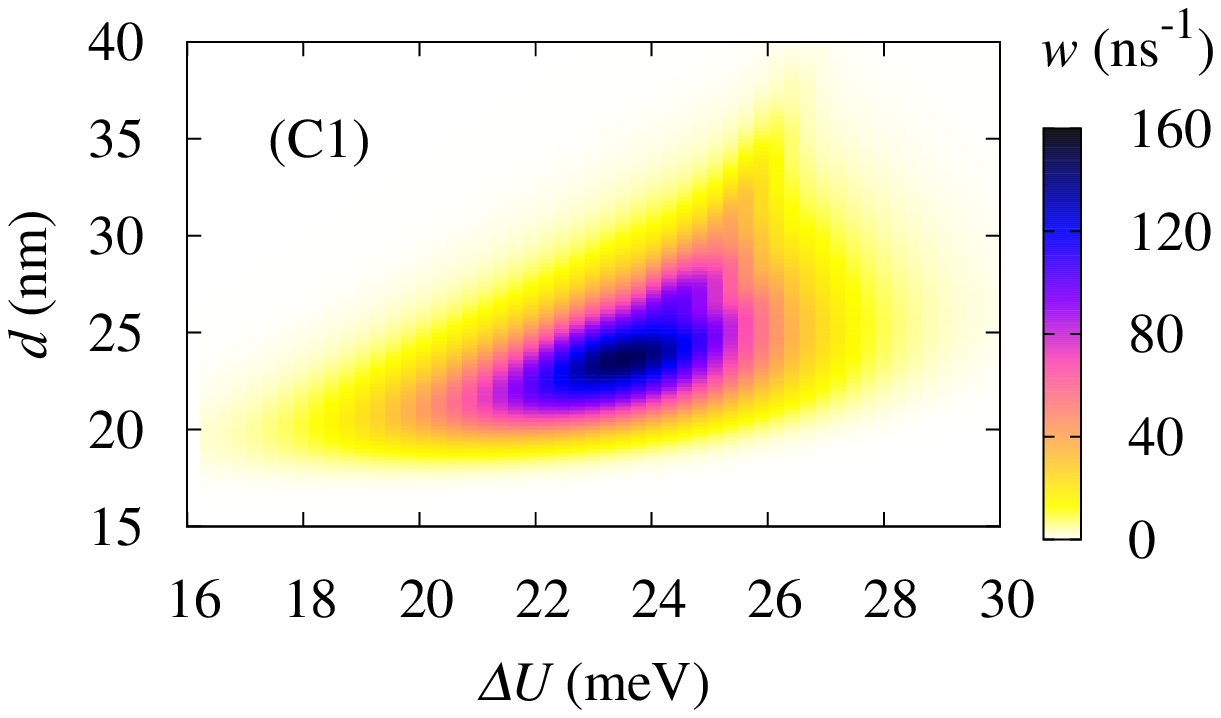}}\hfill
{\includegraphics*[width=.33\textwidth,height=4.0cm]{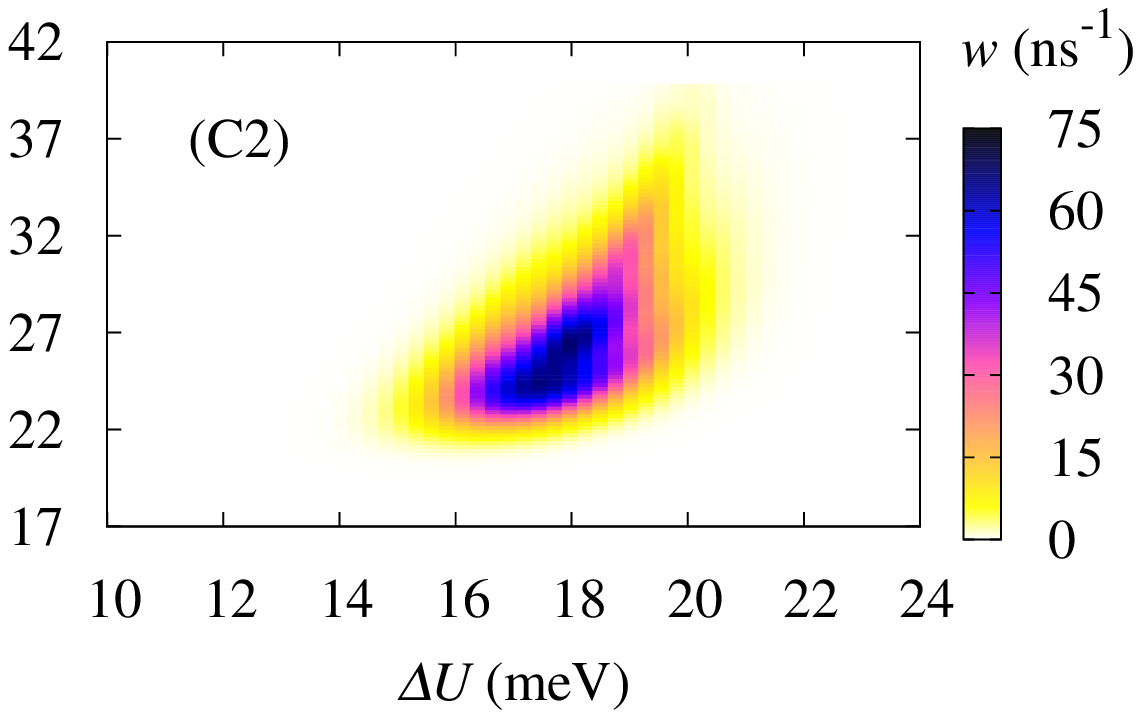}}\hfill
{\includegraphics*[width=.33\textwidth,height=4.0cm]{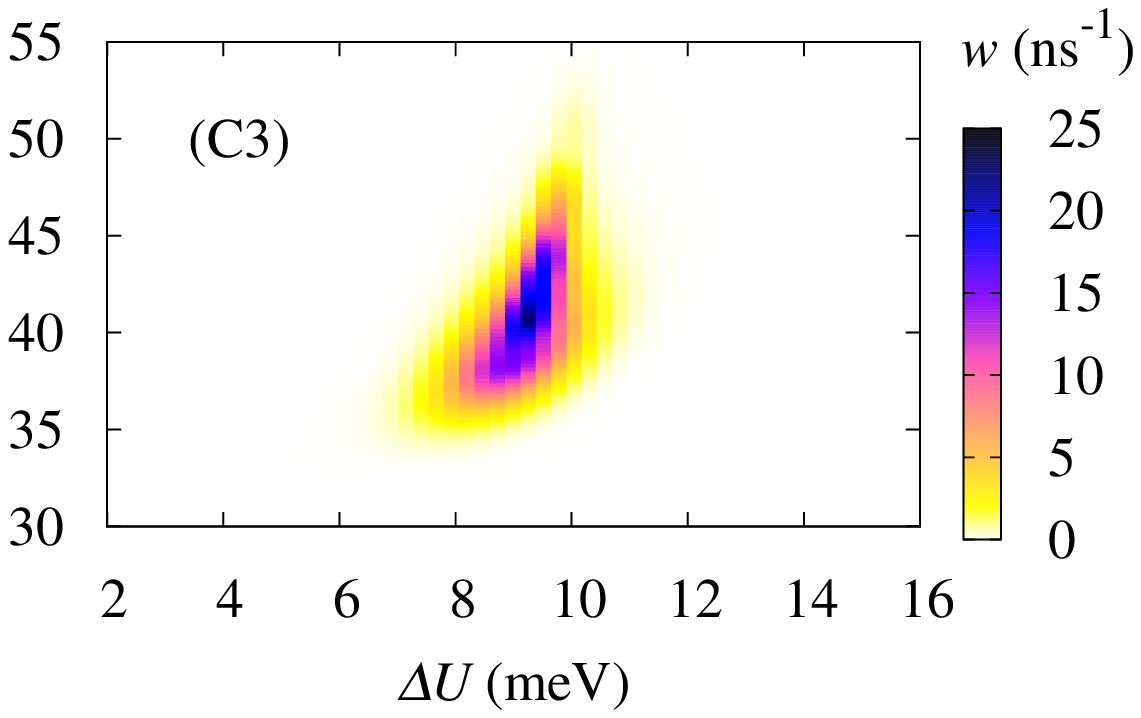}}
\caption{Top: Schematic plots of the three considered QD sizes.
(A1), (A2), and (A3) Tunneling rate assisted by phonons via deformation potential coupling 
as a function of the offset $\Delta U$ and distance $d$ between QDs.
(B1), (B2), and (B3) Tunneling rate assisted by phonons via piezoelectric coupling.
(C1), (C2), and (C3) Total tunneling rate.} \label{fig:set}
\end{figure*}

First, the relaxation rates due to interaction with phonons via deformation potential are considered.
The resulting tunneling rates are plotted in Figs.~\ref{fig:set}(A1)-(A3) for three different quantum dot sizes
(see the sketches on top) as functions of the offset $\Delta U$
and the distance between QDs $d$.
For all three configurations of QDMs, the rates are not symmetric with respect to the offset 
(in contrast to the QDM doped with only one electron \cite{grodecka08}), 
which results from the Coulomb interaction between electrons.
The maximal value of the rate is shifted towards larger offsets with growing distance.
For QDs close to each other, the splitting energy is larger than the acoustic phonon energies,
which leads to low relaxation rates. 
Next, in a certain range of distances, the energy splittings are comparable with phonon energies
and tunneling processes are very efficient, which leads to the high relaxation rates.
For large distances, the relaxation rates decrease again, which results from the decreasing overlap between the two corresponding electron singlet wave functions. The rates are low for small offsets, where the Coulomb interaction prevents the electron from tunneling, and for large offsets, where the energy gap between levels is very large.
The values of the tunneling rate are smaller for larger structures 
since the carrier-phonon interaction is, in general, weaker.
The relaxation processes cover smaller parameter ranges for larger QDs since such structures need larger distances and the range of energies fitting to the acoustic phonon energies is smaller.

In Figs.~\ref{fig:set}(B1)-(B3), the tunneling rates resulting from piezoelectric coupling are plotted. They cover smaller parameter ranges than rates due to deformation potential interaction
since their spectral density covers only low frequencies. 
It is interesting to note that in the case of small quantum dots, this contribution is much smaller then 
the one resulting from deformation, whereas for larger QDs, the values of relaxation rates due to both phonon
interactions are comparable. This results from the fact that the tunneling of an electron leads to larger change of charge distribution in larger structures (larger distances between QDs) 
and hence to larger piezoelectric effects.
Therefore, this coupling is of great importance in double quantum dot structures  in contrast to the phonon response after optical excitations in single QDs, where the piezoelectric coupling 
is in many cases negligible \cite{krummheuer02,forstner03,grodecka07}.

The total rates of tunneling involving all phonon modes are shown 
in Figs.~\ref{fig:set}(C1)-(C3). For the smallest considered QDM, 
the tunneling rates are high for offsets between $\Delta U = 17$ and $\Delta U = 29$~meV 
and for distances from $d = 18$ to $d = 38$~nm. They reach the maximal values of $\sim 160$~ns$^{-1}$ 
for $\Delta U \sim 23$~meV and $d \sim 23$~nm.
For symmetric QDs, the high rates cover smaller values and range of the offset (from $\Delta U = 14.5$ to $\Delta U = 21$~meV) and similar range but larger values of the distance (from $d = 20$ 
to $d = 40$~nm). The maximum of $75$~ns$^{-1}$ is reached for $\Delta U \sim 18$~meV and $d \sim 26$~nm.
For large QDs, the high relaxation rates up to $25$~ns$^{-1}$ cover the offset range
from $\Delta U = 7$ to $\Delta U = 11$~meV and the distance range from $d = 34$ to $d = 50$~nm.

\begin{figure}[h]%
\includegraphics*[width=0.45\textwidth,height=5.5cm]{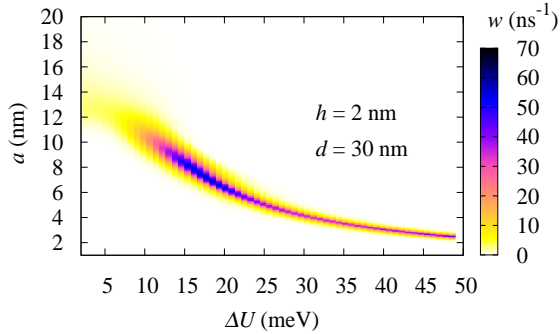}
\caption{Tunneling rate as a function of the offset $\Delta U$ and the size of the QD 
in $x$ and $y$ directions $a$ ($h=2$~nm and $d = 30$~nm).} \label{fig:a}
\end{figure}

We have shown the tunneling rates for three different sizes of QDs. 
To have a better insight in the dependence of the relaxation rates on the QD sizes,
we present in Fig.~\ref{fig:a} the rate as a function of the offset 
and of the lateral QD size $a$. The rates are calculated for fixed QD size
in the growth direction, $h = 2$~nm, and the distance between QDs $d=30$~nm at $T=0$~K.
The maximal values of the tunneling rate are shifted towards smaller offsets with growing QD size.
This results from the fact that for larger structures, the energy differences are smaller
and the Coulomb interaction is weaker, so one does not need a large QD offsets for phonon-assisted tunneling
taking place in the $x$ direction.

\begin{figure}[h]%
\includegraphics*[width=0.45\textwidth,height=5.5cm]{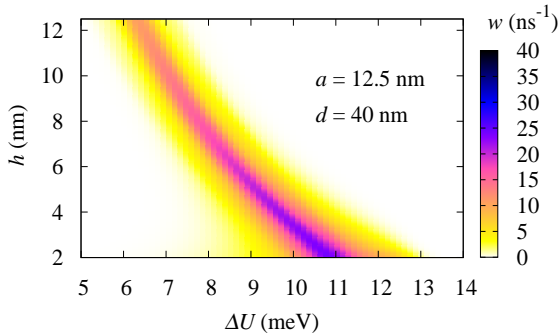}
\caption{Tunneling rate as a function of the offset $\Delta U$ and the size of the QD in $z$ direction $h$
($a=12.5$~nm and $d = 40$~nm).} \label{fig:h}
\end{figure}

Next, we study the relaxation rate as a function of the QD size in the growth direction $h$, 
which is presented in Fig.~\ref{fig:h}. Here, the dependence is simpler than for the different QD sizes
in the $x$ direction, where the electron tunneling occurs. The relaxation rates have higher values for
smaller $h$ and cover a wider sector of the offset. The electron-phonon interaction is stronger for smaller
structures, since the frequency range of phonon spectral density is larger with a cutoff at $\omega_s \sim c_s/h$.

\begin{figure}[h]%
\includegraphics*[width=0.45\textwidth,height=5.5cm]{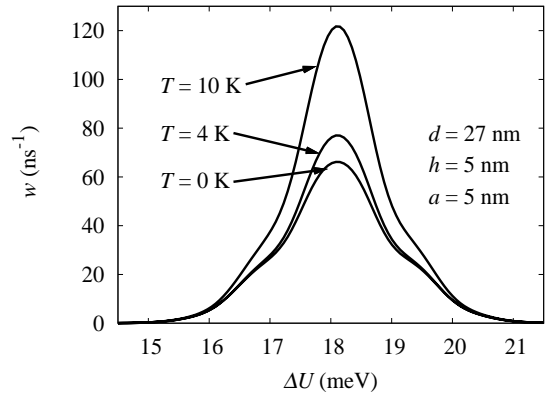}
\caption{Tunneling rate as a function of the offset $\Delta U$ at three different temperatures.} \label{fig:temp}
\end{figure}

Up to now, all the presented results were calculated at $T=0$~K. 
In Fig.~\ref{fig:temp}, the tunneling rates at three different temperatures are shown.
One can see the strong temperature dependence: already at $T=10$~K, the rates are higher by factor $2$
than at zero temperature. This is a direct consequence of the scaling behavior of the Bose distribution function.

\section{Conclusion}

We have studied phonon-assisted tunneling rates in semiconductor quantum dot molecules. For structures doped 
with two electrons, the singlet-singlet relaxation channel was considered. 
The carrier-phonon interaction due to deformation potential as well as piezoelectric coupling was studied in detail, and the importance of the piezoelectric effects in QDM was indicated.
The dependence of tunneling rates on the QDM parameters like the sizes, the offset and the distance between constituent QDs, as well as on temperature was analyzed. 

The phonon-assisted tunneling rates reach high values, so comparing with the spin coherence times on a milisecond timescale \cite{kroutvar04} the process is up to several orders of magnitude faster. The relaxation rate values are also comparable with relaxation times in a single QDs \cite{zibik04}.
This shows, that one cannot neglect phonon-assisted tunneling processes with such time scales.

\begin{acknowledgement}
A. G. and J. F. acknowledge support from the Emmy Noether Program 
of the Deutsche Forschungsgemeinschaft (Grant No. FO 637/1-1). 
\end{acknowledgement}

\bibliographystyle{pss}
\bibliography{abbr,quantum}

\end{document}